\documentstyle[prb,aps,epsfig,multicol]{revtex}
\begin{document}
\draft
\sloppy
\title
{Lattice bosons in a quasi-disordered environment: The effects 
of next-nearest-neighbor hopping on localization 
and Bose-Einstein condensation}
\author{R. Ramakumar$^{1}$, A. N. Das$^{2}$, and S. Sil$^{3}$}
\address{$^{1}$Department of Physics and Astrophysics, 
University of Delhi,Delhi-110007, India} 
\address{$^{2}$Saha Institute of Nuclear Physics,
1/AF Bidhannagar, Kolkata-700064, India} 
\address{$^{3}$Department of Physics, Visva Bharati,
Santiniketan-731235, India}
\date{9 January 2014}
\maketitle
\begin{abstract}
We present a theoretical study of the effects of the next-nearest-neighbor (NNN)
hopping ($t_2$) on the properties of non-interacting bosons in optical lattices
in the presence of an Aubry-Andr\'{e} quasi-disorder. First we investigate,
employing exact diagonalization, the effects of $t_2$ on the localization 
properties of a single boson. The localization is monitored using an 
entanglement measure as well as with inverse participation ratio. We find that 
the sign of $t_2$ has a 
significant influence on the localization effects.
We also provide analytical results in support of the trends found in 
the localization behavior. Further, we extend these results including the
effects of a harmonic potential which obtains in experiments. Next, we study the 
effects of $t_2$ on Bose-Einstein condensation. We find that, a positive
$t_2$ strongly enhances the low temperature thermal depletion of the
condensate while a negative $t_2$ reduces it. It is also found that, for
a fixed  temperature, increasing the quasi-disorder strength reduces the condensate
fraction in the extended regime while enhancing it in the localized regime.
We also investigate the effects of boundary conditions and that of the 
phase of the AA potential on the condensate. These are found to have
significant effects on the condensate fraction in the localization transition
region.
\end{abstract}
\maketitle
\section{Introduction}
\label{sec1}
Cold atoms in harmonic traps and optical lattices continue to be 
an important controllable system for investigations into various properties of
condensed matter. A case in point is the direct observation of Anderson
localization\cite{anderson} of matter waves by several experimental groups
\cite{billy,kondov,jendrzejewski} in recent years. That the Anderson 
localization is a strongly dimension dependant phenomenon was 
recognized early on\cite{mott}. An infinitesimal random disorder localizes all the 
single particle states in two and lower dimensions\cite{mott,abrahams}
(see also the note in Ref. 7). This also rules out the development of 
mobility edges in two and lower
dimensions. However, if the disorder distribution is deterministic, it is
possible to have extended states if the quasi-disorder strength ($\lambda$) is below
a critical value ($\lambda_c$), as is found in the one-dimensional 
Aubry-Andr\'{e} (AA) model\cite{aubryandre}. Recently, localization properties
of non-interacting bosons loaded into a one-dimensional optical lattice with an 
AA quasi-disorder potential were experimentally investigated 
by the LENS group\cite{roati}. 
Detailed theoretical studies of the AA model have been carried out both in 
the past\cite{sokoloff,suslov,soukoulis,dy,thouless,thouless2,kohmoto,kohmoto2,weire,ostlund,wiecko,wilkinson,chao,zdetsis,ingold,roth,aulbach} and 
in the recent 
times\cite{roux,deng,ros,modugno,larcher,adhikari,deng2,cestari,albert,larcher2,cestari2,rkrand}
(for recent reviews see Refs. \onlinecite{modugno2,spalencia,shapiro}).
In the AA model, where the hopping is restricted to between nearest neighbors (NN), 
all the single particle states remain extended for 
$\lambda$ below $\lambda_c$ and become localized above it, which implies, 
again, the absence of mobility edges. Following a line of
research into the effects of longer range hopping on the
localization properties\cite{riklund,rodriguez,xiong,malyshev,demoura}, 
recently it was discovered that the hopping beyond the NN
can lead to the development of mobility edges in what may be called
extended AA models\cite{boers,biddle1,biddle2,biddle3}.
Considering the fact that the AA model has already been realized in
experiments and that the experimental investigations of various
aspects of the localization phenomena using cold atoms
is ongoing\cite{deissler1,deissler2,jendrzejewski2,allard}, further
theoretical studies of the beyond-NN-effects on the localization and the Bose
condensation in extended AA models are certainly of current interest.
The purpose of this paper is to present some interesting results obtained
in such an investigation. 
\par
In this paper, we study the effects of the NNN hopping on some properties
of bosons in optical lattices with AA quasi-disorder without and
with confining harmonic potentials.
In the first part of this paper, we investigate the effects of 
$t_2$ on the localization properties of a single boson employing
the exact diagonalization method. We consider a single boson
moving with NN and NNN hoppings in a one-dimensional (1d) optical 
lattice with AA quasi-disorder. We will show that the sign
of $t_2$ plays an important role on the localization.
The numerical results obtained are supplemented with analytical
results for the energy dependence and the $t_2$ dependence of
the critical disorder strength required for the localization 
transition. We also compare some of the results obtained using
periodic boundary conditions to those obtained employing
open boundary conditions\cite{note} after including a harmonic potential,
which is usually present in the experiments. In the second part
of this paper, we consider a many boson system and study the effects
of $t_2$ on the Bose-Einstein condensation. The condensate fraction
is found to show a significant dependence on $t_2$ and the quasi-disorder
strength. We also show that the phase of the AA potential has a significant
effect on the condensate fraction.
Since we are dealing with a quasi-disordered system, it
is appropriate for us to place this work in the larger field of
the studies on the effects of random disorder on Bose-Einstein
condensates. There have been extensive studies on the random
disorder effects on interacting continuum 
bosons\cite{haung,giorgini,lopatin,astrakharchik,kobayashi,zobay,timmer,yukalov,yukalov2,falco,pilati}.
These investigations found that the condensate fraction decreases with
increasing disorder strength. Further, recent studies on non-interacting
lattice bosons\cite{dellanna,dellanna2} reached the conclusion that the 
Bose condensation temperature slightly enhances with disorder for large filling
whereas it reduces for small filling. 
Extensive studies have also been conducted on
interacting disordered lattice bosons employing the Bose-Hubbard model
and its variants\cite{note2,fisher,soyler,lin,pisarski,kruger,iyer}. 
On the experimental front, it has been reported that
the random disorder reduces the condensate fraction of lattice bosons
in a harmonic trap\cite{white}.
\par
The rest of this paper is organized as follows. The numerical and the analytical
studies on the single particle localization properties are presented in Sec. II. 
The studies on the effect of $t_2$ on the Bose-Einstein condensation and the effects
of the different phases of the AA potential are presented in Sec. III. 
The conclusions are given in Sec. IV.
\section{The effects of the NNN hopping on the localization properties}
\label{sect2}
In the first part of this section we study a lattice boson in the AA disorder
potential. The Hamiltonian of this system is:
\begin{equation}
H=\sum_{i,\delta}\left(t_{1}c^{\dag}_{i}c_{i+\delta}
  +t_{2}c^{\dag}_{i}c_{i+2\delta}\right)
  +\lambda\sum_{i} cos(2\pi qi)c^{\dag}_{i}c_{i},
\end{equation}
where $i$ is a  site index in a one-dimensional
optical lattice with a lattice constant $a$,
$t_1$ and $t_2$ are the NN and the NNN hopping matrix elements,
$c^{\dag}_{i}$ is a creation operator for a boson at site $i$,
$\bf{\delta}$ is the locator of a NN site,
$\lambda$ the strength of the AA potential, and
$q$ = ($\sqrt{5}$+1)/2 is the incommensurability parameter.
Here $t_1$, $t_2$ and  $\lambda$ have energy units.
All the energies in this paper are measured in units of $|t_1|$.
We note here that Biddle {\em et. al.}\cite{biddle2,biddle3}
have shown that, for positive $t_1$ and $t_2$, the $\lambda_c$ decreases with
$t_2$ for low energy states while it increases for high energy
states. In our studies we have considered both negative and
positive $t_2$ values. In this paper, the results presented
for localization studies are for positive $t_1$ unless stated otherwise.
We numerically diagonalize $H$ to obtain its eigenenergies and
eigenfunctions. All our results presented in this paper are
obtained considering a lattice of 233 sites. We monitor the localization
properties by calculating the Shannon
entropy which is a measure of the quantum entanglement\cite{stephan}.
The Shannon entropy for the ground state is given by
\begin{equation}
\displaystyle
S\,=\,-\sum_{i}p_{i}$ log$_{2}\,p_{i},
\end{equation}
where,
\begin{equation}
p_{i}=\left|<i|\psi_G>\right|^{2}=|a_{i}|^{2}
\end{equation}
in which the $a_i$ is the $i$th site amplitude of the ground state wave-function
\begin{equation}
\displaystyle
\left|\psi_G\right> = \sum_{i}a_{i}\left|i\right>.
\end{equation}
We also supplement these results with calculations of the Inverse Participation
Ratio (IPR). The IPR is defined by
\begin{equation}
\displaystyle
IPR =\frac{\sum_{i}p_{i}^{2}}{\left(\sum_{i}p_{i}\right)^{2}}.
\end{equation}
\par
In Fig. 1, we have shown the variations of the S and the IPR of the ground state
as a function of the quasi-disorder strength for various values of $t_2$. The
transition from the extended to the localized state is signaled by a large drop in
$S$ and a large rise in IPR. In the absence of the NNN hopping, the transition
occurs at a critical disorder strength $\lambda_c \approx 2$. 
It is seen that $\lambda_c$ changes as $t_2$ is introduced and the sign of $t_2$ 
has a significant effect on the localization transition. The $\lambda_c$ for 
the ground state decreases with $|t_2|$ for $t_2 >0$ and increases with $|t_2|$
 for $t_2$ negative. In fig. 1 we have also presented the results for $t_1 =-1.0$ 
and $t_2 = -0.2$. We have seen that the results do not depend on the sign of 
$t_1$ \cite{schro}.
\par
Next we have studied the energy dependence of the $\lambda_c$.
We generalize the definitions of S and IPR to excited states by replacing
$\left|\psi_G\right>$ with corresponding excited state wave functions in Eq. (3).
We have numerically obtained the $\lambda_c$ for all the 233 eigenstates from the
calculations of the $dS/d\lambda$ as well as the $d(IPR)/d\lambda$ as a function
of $\lambda$ as shown in Fig. 2 for three states. For each eigenstate,
the value of $\lambda$ for which $dS/d\lambda$ is minimum
and $d(IPR)/d\lambda$ is maximum is taken as the $\lambda_c$ of that state
and the corresponding eigenenergy is $E_c$.
We note here that $dS/d\lambda$ is a better 
indicator of the localization transition compared to $d(IPR)/d\lambda$.
In Fig. 3, we have shown the variation of the $\lambda_c$ with
$E_c$ for $t_2$ = -0.1.
The $\lambda_c$ is seen to decrease approximately linearly with increasing energy.
It may be noted that for positive $t_2$ this trend is reverse\cite{biddle3}.
\begin{figure}
\begin{center}
\includegraphics[angle=270.0,width=4in,=4in]{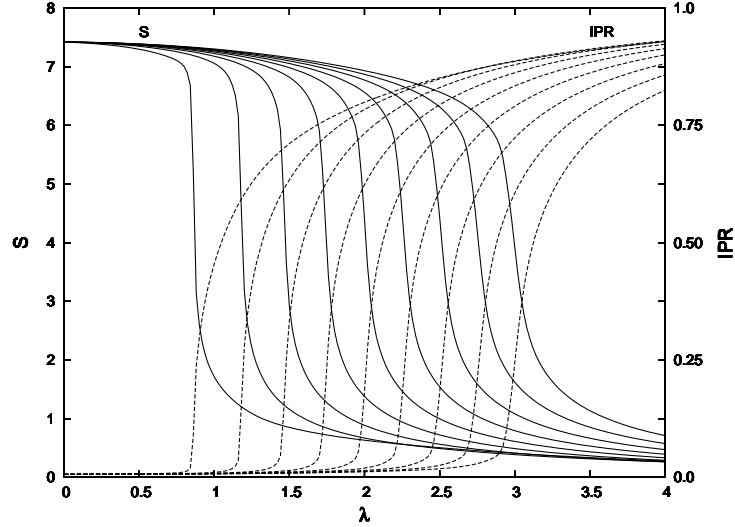}
\caption{  Entanglement (solid) and the IPR (dashes) as a function of the AA 
quasi-disorder potential
strength ($\lambda$) for a closed chain with 233 lattice sites for different
values of $t_2$. From left to right, the curves are for $t_2$ = 0.2, 0.15,
0.1, 0.05, 0, -0.05, -0.1, -0.15, and -0.2, respectively. 
}
\label{fig1}
\end{center}
\end{figure}
\begin{figure}
\begin{center}
\includegraphics[angle=270.0,width=4in,=4in]{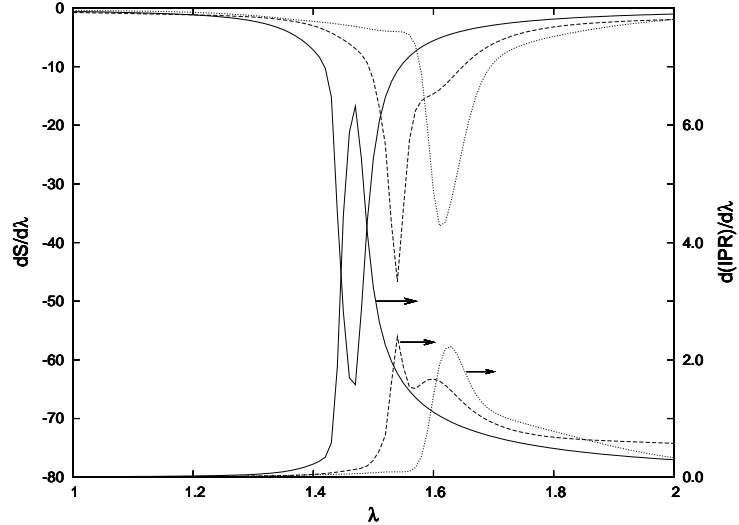}
\caption{  
The variation of $dS/D\lambda$ and $d(IPR)/d\lambda$ with $\lambda$ for
the the ground state (solid), 39th excited state (dash), and the
78th excited state (dot) of a boson in a closed chain of 233 sites.
}
\label{fig2}
\end{center}
\end{figure}
 The numerical results obtained for negative $t_2$ can be understood from
an extension of the analytical calculations given in Ref. \onlinecite{biddle2}.
We start with the following Hamiltonian for an infinite system:
\begin{equation}
\bar{H}=\sum_{n,n^{\prime}}^{\prime}t_{nn^{\prime}}c^{\dag}_{n}c_{n^{\prime}}+
\lambda\sum_{n} cos(2\pi qn)c^{\dag}_{n}c_{n},
\end{equation}
where 
\begin{equation}
t_{nn^{\prime}}=e^{is\pi} te^{-(p+is\pi)|n-n^{\prime}|}
\end{equation}
$n$ and $n^{\prime}$ are the site indices and $s$= 0 or 1. 
Here, the hopping strength decreases exponentially with increasing distance
and our choice (Eq. 7) makes $t_1$ positive 
whereas  $t_2$ is positive or negative  depending on $s=0$ or $1$. 
The relevance of this model in the context of ultracold atoms in optical
lattices has been discussed in Ref. \onlinecite{biddle2}.
The eigenvalue equation is then
\begin{equation}
\left(E+e^{is \pi}t-\lambda cos(2\pi qn)\right)u_n=e^{is\pi}\sum_{n^{\prime}}t
e^{-(p+is\pi)|n-n^{\prime}|}u_{n^{\prime}},
\end{equation}
where $u_n$ is the amplitude of the wave function at site $n$ and $E$
the energy eigenvalue.
Now defining
\begin{equation}
cosh(p_{\circ})=e^{is \pi} \left(\frac{E+e^{is \pi}t}{\lambda}\right)
\end{equation}
which immediately gives
\begin{equation}
sinh(p_{\circ})=e^{is \pi}\frac{\omega}{\lambda}
\end{equation}
with
\begin{equation}
\omega=e^{is \pi}\sqrt{(E+e^{is\pi}t)^{2}-\lambda^{2}},
\end{equation}
the eigenvalue equation becomes
\begin{equation}
\omega T_n(p_{\circ})u_n=e^{is \pi} \sum_{n^{\prime}}
te^{-(p+is\pi)|n-n^{\prime}|}u_{n^{\prime}},
\end{equation}
where
\begin{equation}
T_n(p_{\circ})=\frac{cosh(p_{\circ})-e^{is \pi}cos(2\pi qn)}{sinh(p_{\circ})}.
\end{equation}
The dual to the preceding eigenvalue equation is obtained by multiplying
both sides of it
with $exp(i2\pi mnq)$ and summing over $m$, an integer. The dual is obtained as
\begin{equation}
\omega T_n(p)\tilde{u}_n =e^{is \pi}\sum_{n^{\prime}}
te^{-(p_{\circ}+is\pi)|n-n^{\prime}|}
\tilde{u}_{n^{\prime}},
\end{equation}
where
\begin{equation}
\displaystyle
\tilde{u}_m=\sum_{n}T_{n}(p_{\circ})e^{i2\pi mnq}u_{n}.
\end{equation}
The Eqs. (12) and (14) become self dual for $p$ = $p_{\circ}$.
At the self dual point, following Eq. (9), one gets
\begin{equation}
cosh(p)=e^{is \pi}\left(\frac{E+e^{is \pi}t}{\lambda}\right).
\end{equation}
Noting that $t_1 = t e^{-p}$ and $t_2 = e^{is \pi}t e^{-2p}$, one obtains from
the preceding equation an equation for $\lambda_c$ as
\begin{equation}
\lambda_c =\frac{2t_1 + 2e^{is \pi}E e^{-p}}{1+e^{-2p}} =
\frac{2t_1 + 2E\,(t_2/t_1)}{1+(t_2/t_1)^{2}}.
\end{equation}
When $t_2$ = $0$, the critical disorder strength $\lambda_c$
is energy independent and is equal to $2t_1$, a result obtained
for the original AA model. When $t_2 \neq$ $0$ and energy ($E$) is fixed,
the change in $\lambda_c$ is proportional to $t_2$ provided $t_2/t_1$ is
small. For a fixed value of $t_2$, $\lambda_c$ increases or decreases
linearly with the energy eigenvalue $E$ depending on $t_2$ being
positive or negative.
In Fig. 3, the solid line is obtained by using Eq. (17) for $t_2/t_1$= -0.1. 
The numerical results are in reasonable agreement with the analytical results.
Note that the numerical results are obtained considering NN and NNN hoppings only
while the analytical results are derived considering long range hopping which
decays exponentially with distance.
In Fig. 4, we plot the $t_2$ dependence of $\lambda_c$ for the ground state. 
We find that an increase in  $|t_2|$ for negative $t_2$ increases 
the $\lambda_c$ while
increasing positive $t_2$ reduces $\lambda_c$ and that it decreases almost 
linearly with increasing $t_2$. 
\par
   From now on, keeping the cold atom experiments in mind, we consider 
the effects of a harmonic confining potential. The Hamiltonian
of the system is then 
\begin{equation}
\tilde{H}=H+\sum_{i}ki^{2}c^{\dag}_{i}c_{i},
\end{equation}
where $k$, which has an energy unit, is the strength of the harmonic
confining potential. 
In the presence of harmonic potential, we use open boundary conditions 
and measure the position coordinate of a lattice site from the center
of the harmonic trap. In such cases the effect of the 
second nearest-neighbor hopping on the entanglement is shown in Fig. 5.
In this figure one finds a small region of $\lambda$ where the 
entanglement ($S$) increases with $\lambda$, in contrast to the usual 
behavior of suppression of $S$ with increasing $\lambda$. Then $S$ reaches
a peak and falls abruptly with increasing $\lambda$ signifying localization 
of the wave function.      
%
The small enhancement of $S$ preceding the localization transition 
results from a competition between the AA potential trying to
move the boson away from the center of the trap and the confining potential
trying to bring it to the center of the trap, as explained
in our earlier paper (Ref. \onlinecite{rkrand}). 
We note here that the disorder induced enhancement of $S$  becomes prominent 
and the peak becomes sharper with increasing positive $t_2$, whereas the 
peak becomes broader for negative values of $t_2$. 
In Fig. 6 we have plotted the $\lambda_c$ for the ground state 
as a function of $t_2$ for different values of the strength ($k$) of 
the harmonic potential.  $\lambda_c$ decreases with increasing $t_2$ 
and the curves are almost linear as similar to that for a closed chain. 
For a fixed $t_2$ the $\lambda_c$ increases with increasing $k$.   
The physical origin of this effect is 
in the harmonic potential opposing the disorder trying to localize 
the boson away from the trap center.
\begin{figure}
\begin{center}
\includegraphics[angle=270.0,width=4in,totalheight=3.50in]{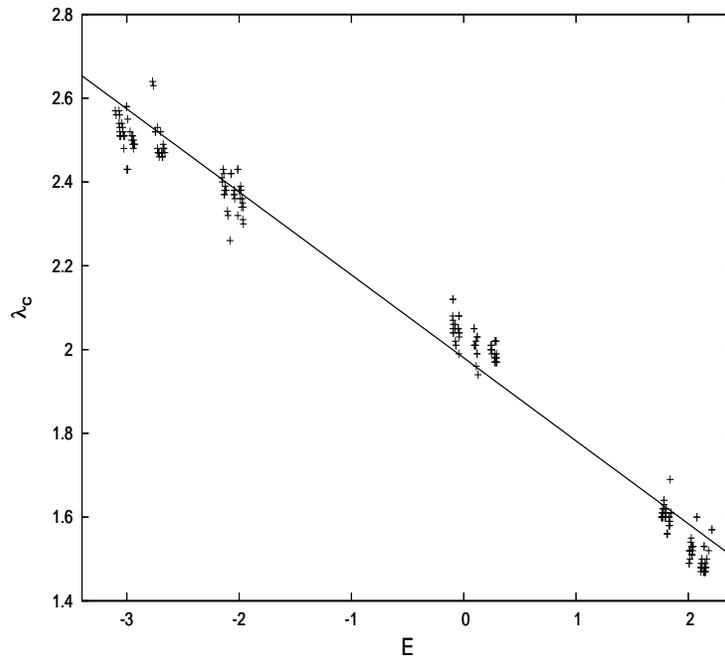}
\caption[]
{Plot of $\lambda_c$ vs E for $t_2$ = -0.1. The plus signs represent 
numerically determined $\lambda_c$ and energy eigenvalues for different 
eigenstates for a closed chain of 233 lattice sites and the solid 
line is the approximate analytical result. 
}
\label{fig3}
\end{center}
\end{figure}
\begin{figure}
\begin{center}
\includegraphics[angle=270.0,width=4in,totalheight=2.5in]{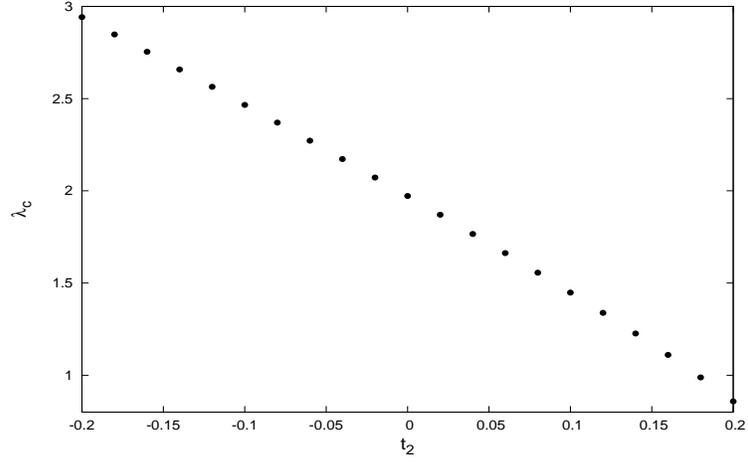}
\caption[]
{The variation of the $\lambda_c$ with $t_2$ for a closed chain 
of 233 lattice sites.}
\label{fig4}
\end{center}
\end{figure}
\begin{figure}
\begin{center}
\includegraphics[angle=270.0,width=4in,totalheight=2.5in]{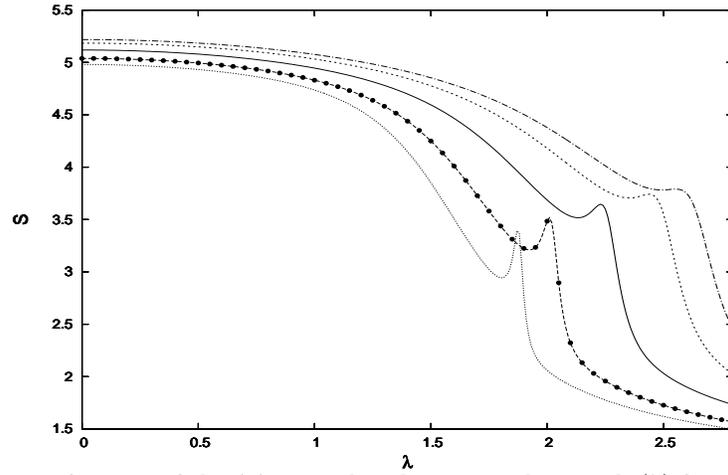}
\caption[]
{
Entanglement (S) as a function of the AA quasi-disorder potential
strength ($\lambda$) for an open chain with 233 lattice sites in
a harmonic trap of $k$ = 0.00005 for different
values of $t_2$. From left to right, the curves are for $t_2$ = 0.08,
0.05, 0, -0.05, -0.08, respectively.
The AA potential is placed symmetrically about the center of
the harmonic trap with phase factor $\phi$ = 0 (see Eq. 22).
The solid circles represent the results for $t_1$ = -1.0 and $t_2$ = 0.05.}
\label{fig5}
\end{center}
\end{figure}
\begin{figure}
\begin{center}
\includegraphics[angle=270.0,width=4in,totalheight=2.75in]{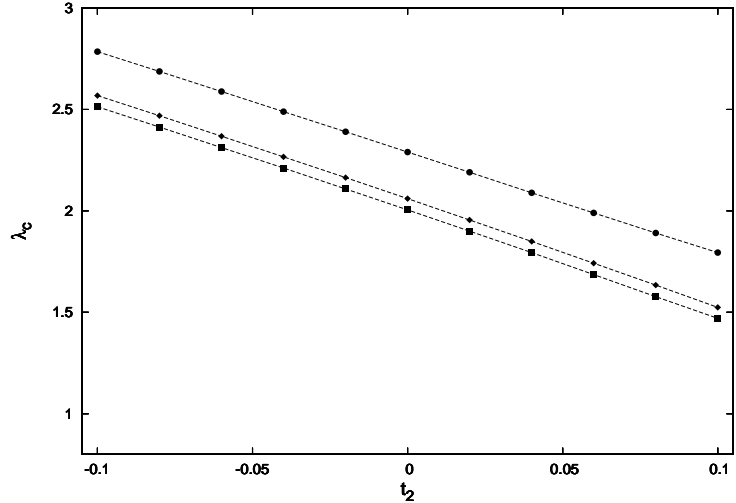}
\caption[]
{
The variation of $\lambda_c$ with $t_2$ for an open chain
of 233 lattice sites in harmonic traps: $k$ = 0 (squares),
$k$ = 0.00001 (diamonds), and $k$ =0.00005 (circles). 
The AA potential is placed symmetrically about the center of
the harmonic trap with phase factor $\phi$ = 0.}
\label{fig6}
\end{center}
\end{figure}
\section{The effects of NNN hopping on the Bose-Einstein condensation}
\label{sec3}
In this section we first study the effect of $t_2$ on the
Bose condensate fraction of a collection of bosons in
a finite one-dimensional periodic optical lattice with AA disorder
in a harmonic confining potential. The system Hamiltonian is
\begin{equation}
\tilde{\tilde{H}}=\tilde{H}-\mu\sum_{i}c^{\dag}_{i}c_{i},
\end{equation}
where $\mu$ is the chemical potential.
In terms of the single particle energy levels ($E_i$)obtained by numerically
diagonalizing the Hamiltonian given in Eq. (18), the boson number equation is
\begin{equation}
N\,=\,\sum_{i=0}^{m}N(E_{i}),
\end{equation}
where $E_0$ and $E_m$ are the lowest and the highest energy levels, and
\begin{equation}
N(E_{i})=\frac{1}{e^{\beta\,(E_{i}-\mu)}-1}
\end{equation}
in which $\beta\,=\,1/k_{B}T$ with $k_{B}$ the Boltzmann constant
and $T$ the temperature. We first determine the chemical potential and
then the boson populations in various energy levels using the
boson number equation. 
\par
The temperature dependence of the condensate fraction for various
values of the quasi-disorder strength for fixed $t_2$ values are
shown in Fig. 7. In the low temperature regime, increasing
$\lambda$ is seen to suppress the condensate fraction. Beyond
this temperature range, increasing $\lambda$ leads to a reduction
of the $N_0/N$ until $\lambda$ = $\lambda_c$ and then to an 
enhancement for $\lambda \, >\, \lambda_c$.
The effect of $t_2$ on the condensate fraction for fixed
value of the quasi-disorder strength is shown in Fig. 8.
It is clear that a negative $t_2$ 
enhances the $N_0/N$ while a positive $t_2$ reduces it.
It is also seen that the effect of a positive $t_2$
is more prominent in comparison with a negative $t_2$. 
\begin{figure}
\begin{center}
\includegraphics[angle=270.0,width=4in,totalheight=5in]{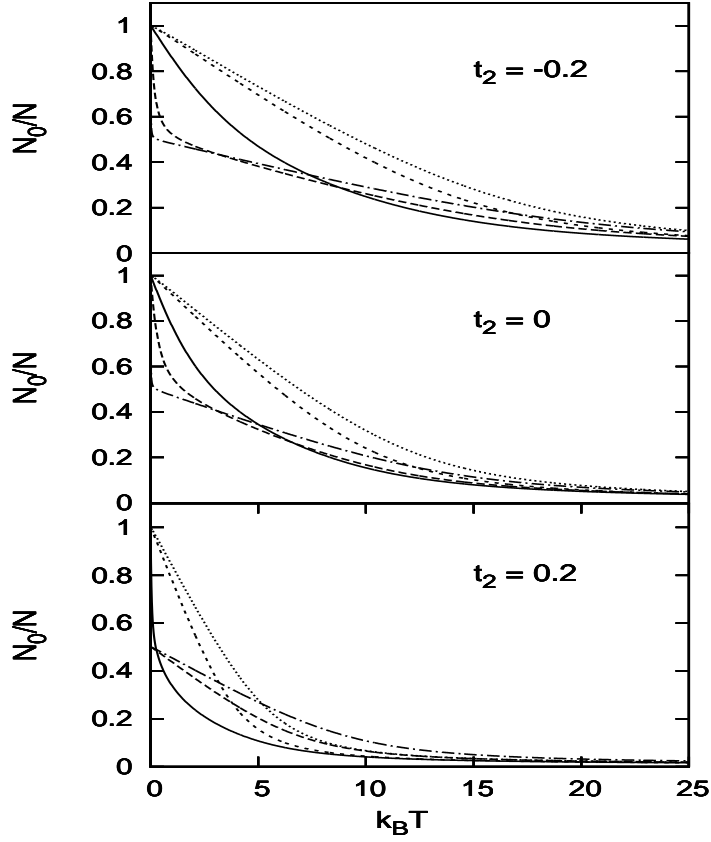}
\caption[]
{
The variation of the condensate fraction ($N_0/N$) with temperature for
10000 bosons in an open chain with 233 sites in a harmonic trap with $k$ = 0.00001 for
different strengths of the AA potential and different values of $t_2$.
Top panel: $\lambda$ = 0 (dots), 2.0 (short dashes), 3.05 (solid),
3.5 (long dash), and 4.0 (dash-dot). 
Middle panel: $\lambda$ = 0 (dots), 1.5 (short dash), 2.06 (solid), 
2.2 (long dash), and 2.5 (dash dot). 
Bottom panel: 0 (dots), 0.8 (short dash), 0.94 (solid),
1.2 (long dash), and 1.5 (dash dot).
The solid line is for $\lambda$ = $\lambda_c$.
The AA potential is placed symmetrically about the center of
the harmonic trap with phase factor $\phi$ = 0.
}
\label{fig7}
\end{center}
\end{figure}
\begin{figure}
\begin{center}
\includegraphics[angle=270.0,width=4in,totalheight=3in]{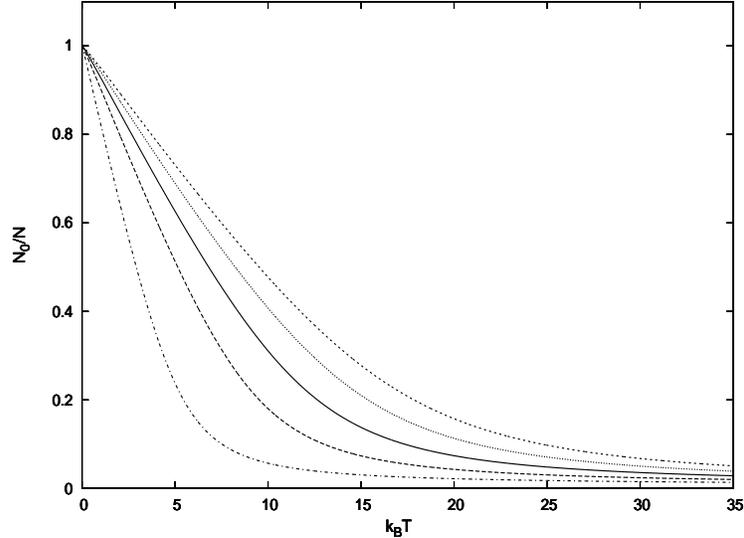}
\caption[]
{
The variation of the condensate fraction with temperature for
10000 bosons in an open chain with 233 sites in a harmonic trap with $k$ = 0.00001 for
$\lambda$ = 0.5 and different values of $t_2$. From top
to bottom, the curves are for $t_2$ = -0.2, -0.1, 0, 0.1, and 0.2,
respectively. 
The AA potential is placed symmetrically about the center of
the harmonic trap with phase factor $\phi$ = 0.
}
\label{fig8}
\end{center}
\end{figure}
\par
All the results presented in the presence of the harmonic trap 
are obtained by placing the AA potential symmetric about the 
center of the trap. To study the effects of
varying the phase factor of the AA potential on the condensate fraction,
we consider the potential at a site $i$ with a phase factor $\phi$ as
\begin{equation}
V(i) =\lambda \, cos[2\pi q i + \phi],
\end{equation}
where the coordinate of the site $i$ is measured from the center of the trap.
For $\phi=0$ the potential is symmetric about the center and has its highest 
value there. This is a unique situation where the AA potential and 
the harmonic trap compete with each other in the localization transition
regime as mentioned previously.   
In the top panel of Fig. 9, we have compared the $\lambda$
dependence of the low temperature $N_0/N$ for $\phi$ = 0 with
two finite values of $\phi$. The phase factor is found to have
a strong influence. The drop in the condensate fraction to 0.5
in the localization transition region occurs only for $\phi$ = 0.
To understand the $\lambda$ dependence of $N_0/N$, we looked
at the energy difference between the ground state and the
first excited state ($\Delta E$), as shown in the lower panel
of the Fig. 9. The physical origin of the variations seen
is in the disorder induced changes in the lower energy
levels of the boson. 
For $\phi$ = 0, the drop to 0.5 occurs because the ground state 
and the first excited state are equally populated since $\Delta E$ 
is extremely small for $\lambda$> $\lambda_c$.  
For non zero values of $\phi$, the $\Delta E$ shows a minimum at 
$\lambda_c$, and then it becomes large enough so that almost all
the bosons are in the ground state for all $\lambda$ except at 
$\lambda_c$ where $N_0/N$ shows a dip. 
In the upper panel of Fig. 10,
we have shown the dependences of $N_0/N$ in the entire of 
range ($0$ to $\pi$) of $\phi$ for two temperatures
for a fixed AA disorder strength. The changes in $N_0/N$ increases
with increasing temperature. The variation of $N_0/N$ closely
tracks the changes in $\Delta E$ given in the lower panel of the
Fig. 10.
\section{Conclusions}
In this paper we investigated the effects of next nearest-neighbor (NNN)
hopping ($t_2$) on the properties of non-interacting bosons in optical lattices
in the presence of an Aubry-Andr\'{e} quasi-disorder. 
In closed chains, for the ground state, a negative $t_2$ enhances the 
critical disorder strength ($\lambda_c$) for the ground state 
required for the localization transition while a positive $t_2$ reduces it.
For high energy states ($E > 0$), the trend is opposite.
These results obtained numerically were complemented with analytical calculations.
We further extended these studies of the single particle localization
including the effects of a harmonic confining potential with open 
boundary conditions which usually obtains in cold atom experiments. 
The harmonic potential was found to 
increase the $\lambda_c$. It was also found that while negative $t_2$
enhances the entanglement (S), the effect of a positive $t_2$ is to reduce it.
Further, a positive $t_2$ enhances the disorder induced enhancement
of the $S$ when the harmonic trap competes with the AA potential in 
the localization transition region.
Next we considered a many boson system and studied the effects of $t_2$
on the Bose condensation. It is found that the thermal depletion of the
condensate is enhanced by a negative $t_2$ while it is reduced
for a positive $t_2$. Finally, we investigated the effects of the
phase of the AA potential on the condensate fraction ($N_0/N$) to find that
it has a very strong effect on the $N_0/N$ for $\lambda \geq \lambda_c$. 
\label{sec4}
\begin{figure}
\begin{center}
\includegraphics[angle=270.0,width=4in,totalheight=3in]{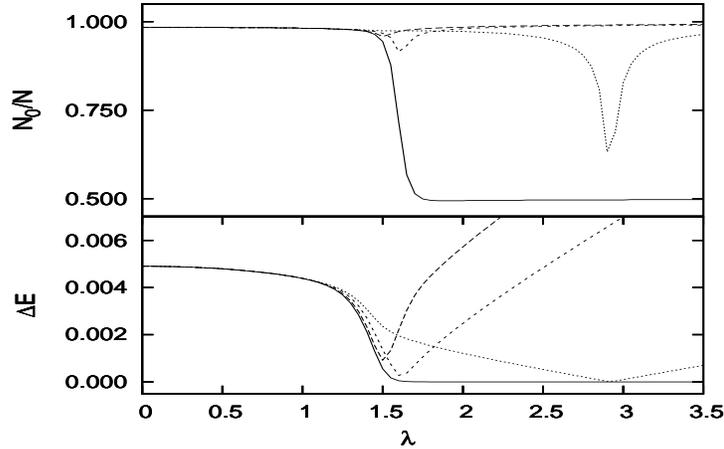}
\caption[]
{
Top panel: The variation of the condensate fraction with $\lambda$
for 10000 bosons in an open open chain with 233 sites in a harmonic trap with $k$ = 0.00001, 
$t_2$ = 0.1 and $k_BT$ = 0.2. The results are for:
$\phi$ = 0 (solid line), 0.05 (long dashes), 0.08 (short dash) and 0.1 (dots). 
The bottom panel shows the corresponding variation of  
$\Delta E$ (= $E_2-E_1$) with $\lambda$.
}
\label{fig9}
\end{center}
\end{figure}
\begin{figure}
\begin{center}
\includegraphics[angle=270.0,width=4in,totalheight=3.75in]{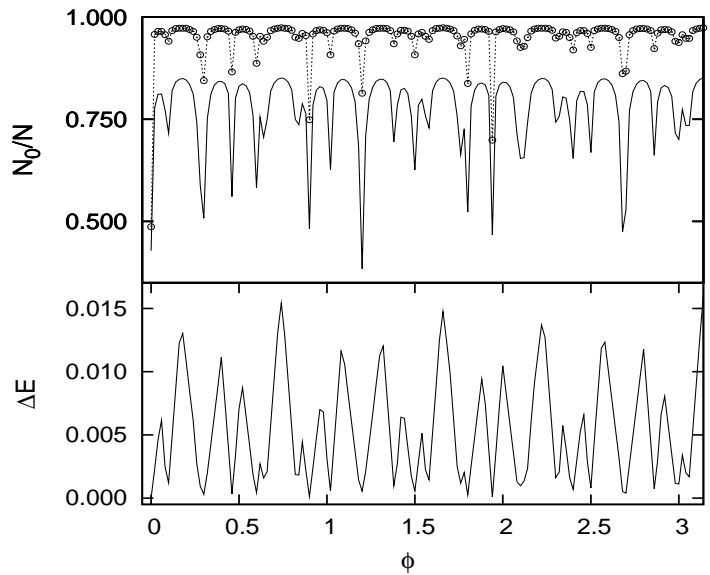}
\caption[]
{
The variation of the condensate fraction (top panel) and 
$\Delta E$  (bottom panel) for
10000 bosons in an open chain with 233 sites in a harmonic trap with $k$ = 0.00001
as a function of the phase factor ($\phi$) for $t_2$ = 0.1 and $\lambda$ =2.0.
The results in the top panel are for: $k_BT$ = 0.4 (open circles with a guiding dotted line)
and 2.0 (solid line). 
}
\label{fig10}
\end{center}
\end{figure}
\acknowledgments
RRK thanks Professor M. K. Sanyal, Director, SINP and
Professor S. N. Karmakar, Head, TCMP Division, SINP
for hospitality at SINP. 


\begin{references}
\bibitem{note}
In the numerical calculations, the closed chain results are
obtained by taking $t_{N_{s},1}=t_{1,N_{s}}=t_1$ and
$t_{N_{s}-1,1}=t_{1,N_{s}-1}=t_{N_{s},2}=t_{2,N_{s}}=t_2$, while for the
open chains these matrix elements are taken as zero, where $N_{s}$ is the number of sites
in the lattice.
\bibitem{note2}
There have been detailed investigations on the effects of disorder on lattice bosons
within the framework of the Bose-Hubbard model since the late eighties 
(Ref. \onlinecite{fisher}). Since our focus is not interacting bosons in this paper, 
we request the reader to consult the recent papers (Refs. \onlinecite{soyler,lin,pisarski,kruger,iyer}), and the references contained therein.
\bibitem{schro} 
In presence of nearest and next nearest neighbor hopping the Schr$\ddot{o}$dinger 
equation reduces to 
$(E-\epsilon_n)u_n=t_1 (u_{n+1} + u_{n-1}) + t_2 (u_{n+2} + u_{n-2})$.
If we transform $u_n$ to $ e^{in \pi} u'_{n}$ then $u'_n$ satisfies similar 
equation with negative $t_1$. Consequently, the eigenvalues 
remain unchanged if the sign of $t_1$ is changed. 
Moreover, $|u_n|^2=|u'_n|^2$ makes the localization behaviour of the system 
unaltered with the change of the sign of $t_1$.
\end{references}
\end{document}